\documentclass[12pt,a4paper]{article}

\usepackage{graphicx}
\usepackage[none]{hyphenat}

\begin{document}

%
%

\title{Vertical midscale ionospheric disturbances caused by surface seismic waves based on Irkutsk chirp ionosonde data in 2011-2016.}

%
%

%
%

\author{O.I. Berngardt, N.P. Perevalova,A.V. Podlesnyi, V.I. Kurkin, G.A. Zherebtsov}



\maketitle









%
%


\begin{abstract}

Based on the Irkutsk fast monostatic chirp ionosonde data
we made a statistical analysis of ionospheric effects for 28 earthquakes
which appeared in 2011-2016 years. These effects are related with surface (Rayleigh) seismic waves far from epicenter. 
The analysis has shown that nine of these earthquakes were accompanied by vertical midscale ionospheric irregularities (multicusp). 
To estimate the ionospheric efficiency of the seismic waves we proposed new index $K_{W}$. 
The index estimates the maximal amplitude of the acoustic shock wave generated by given spatial 
distribution of seismic vibrations and related with maximal spectral power of seismic oscillations. 

Based on the analysis of experimental data we have shown that
earthquake-related multicusp is observed mostly at daytime [07:00-17:00]LST for $K_{W}\ge4.7$.
The observations of intrinsic gravity waves by GPS technique in the epicenter 
vicinity do not show such a daytime dependence.
Based on 24/05/2013 Okhotsk Sea earthquake example, we demonstrated that  deep-focus earthquakes 
can produce strong multicusp far from the epicenter, although do not produce significant 
GPS ionospheric response in the epicenter vicinity.
Three cases of sporadic E bifurcation in far epicentral zone were also detected and discussed.

\end{abstract}

%
%

%


%
%

\section{Introduction}

The investigation of ionospheric effects caused by the Earth's surface
sources is a traditional and effective technique for studies of neutral-ionospheric
interaction. A lot of papers are devoted to theoretical and experimental
aspects of the problem \cite{Blanc_1985,Pokhotelov_1995,Lognonne_1998,Lastovicka_2006,Kiryushkin_2011,Jin_2015}.

One of such phenomena is generation of midscale ionospheric irregularities directly by
shock wave from the supersonic seismic source propagating over the
surface \cite{Maruyama_2011,Liu_2011,Rolland_2011,Kakinami_2013,Maruyama_2014,Berngardt_2015,Maruyama_2016a,Maruyama_2016b,Kherani_2016,Chum_2016,Liu_2016}.

To estimate the seismic disturbance efficiency to the ionosphere is the essential 
problem for planning and interpretation of each single experiment.
The effective generation of intrinsic gravity waves (IGW) in the vicinity of the epicenter
according to GPS-data requires large enough earthquake magnitude \cite{Perevalova_2014}.
But there are no statistical information about the boundary
values for the intensity of seismic (Rayleigh) waves that produce the
vertical midscale ionospheric structures (multicusp).

For the study of the multicusp the vertical ionosondes are 
widely used \cite{Maruyama_2011,Maruyama_2014,Maruyama_2016a,Maruyama_2016b}.
The use of chirp ionosondes for this problem significantly increases the
signal-to-noise ratio and improves the temporal and spatial resolution, which
are the key parameters for such studies. The efficiency of replacement
of the pulse ionosondes by chirp ionosondes for diagnosis of ionospheric profile was demonstrated, for example
in \cite{Harris_2016}.

Baikal region is a seismically active area, so studies of seismic activity effects are constantly being conducted. 
A number of various ionospheric, optical, magnetic, seismic and acoustic instruments are involved into monitoring and case-study tasks \cite{Zherebtsov_2012}.

The aim of this paper is a preliminary statistical analysis of the
emergence of midscale vertical ionospheric irregularities caused by large earthquakes
at a large distance from the epicenter and investigating the effectiveness
of this mechanism.
In order to complete this, we used the 2011-2016 data from the Irkutsk fast
monostatic chirp ionosonde (IFMCI). We also used a variety of GPS-receivers
in the areas near the epicenters of the earthquakes to compare near zone and far zone ionospheric effects.

\section{Data used in the experiments}

\subsection{Chirp ionosonde}

IFMCI (51.81N, 103.078E)  started its continuous operation in 2012, 150 km to the South-West from Irkutsk. 
For operation it uses a continuous chirp-signal with a frequency range from 1.5 MHz to 15 MHz and 10Watts of transmitted power\cite{Podlesnyi_2011, Podlesnyi_2013}. 
The spacing between the receiving and transmitting antenna is approximately 150 meters. Low power and minimal spacing between 
antennas allows transmitter and receiver to operate at the same time. 

The use of modern technologies of digital reception allows us to reach a high flexibility in the use of equipment and get 
a wide dynamic range. The fully digital structure of the ionosonde allows us to precisely control the shape 
of the transmitted signal as well as the impulse response and filtering characteristics of the receivers \cite{Podlesnyi_2011}. We develop and use original techniques 
for filtering out the signals from neighbor public radiostations and significantly improve the quality of the ionosonde data \cite{Podlesnyi_2014}.

The typical frequency sweep speed is about 500 kHz/s, providing 27 second time resolution. From the beginning of 
its continuous operation in 2012 the ionosonde uses 1-minute time resolution typical for modern monitoring techniques \cite{Reinisch_2009,Akchurin_2011,Harris_2012, Kurkin_2014}. 
This allows us to use its data for detailed study of vertical ionospheric disturbances generated by seismic waves.

\subsection{Seismic data sources}

\begin{figure}
\includegraphics[scale=0.65]{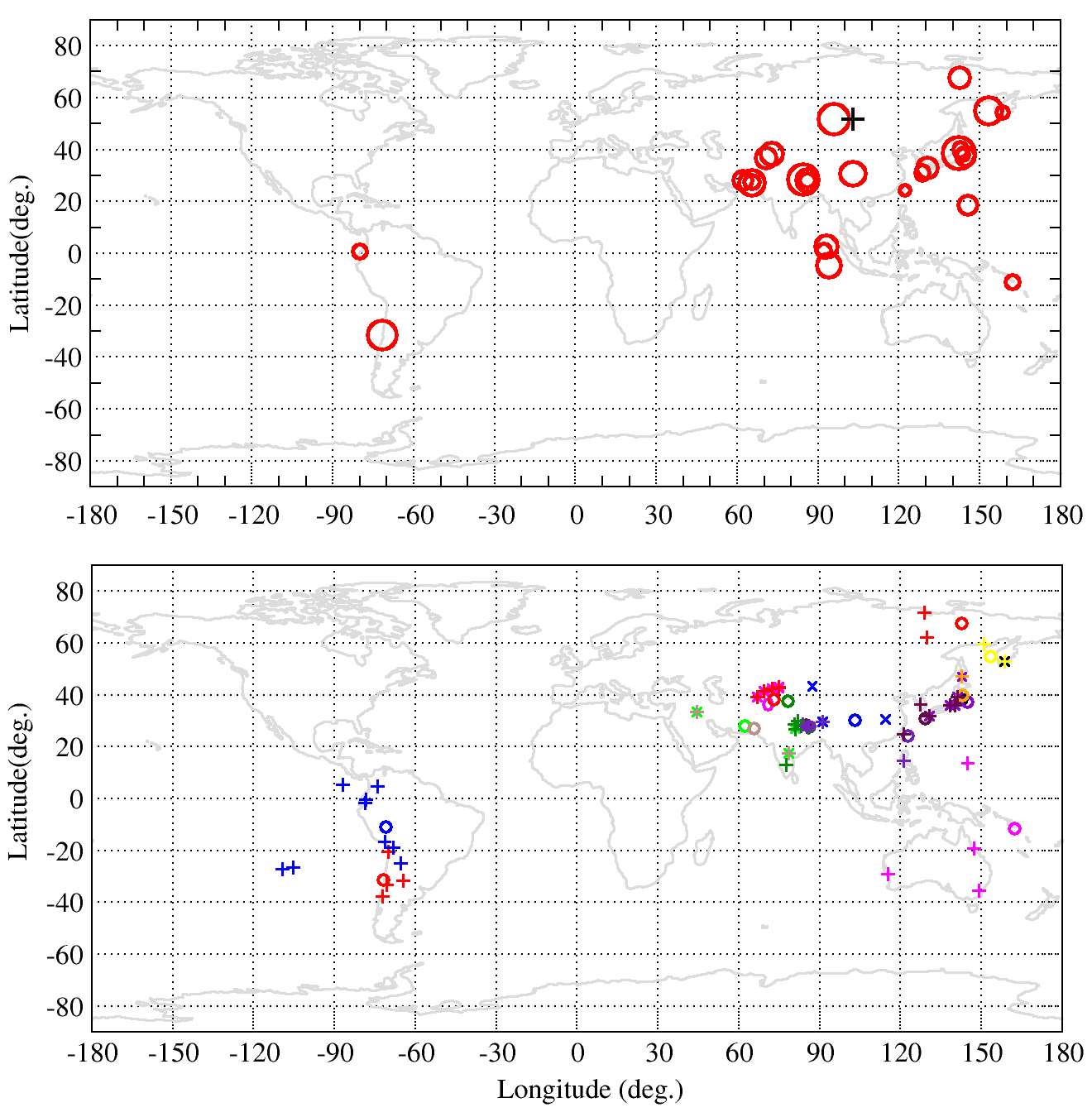}
\caption{
A) Geometry of chirp measurements in 2011-2016. Circles correspond to earthquakes epicenters, their radii corresponds to calculated $K_{A}$. Cross corresponds to measurement location (IFMCI and TLY seismic station).
B) Geometry of GPS measurements in 2013-2015. Circles correspond to earthquakes epicenters, crosses and diagonal crosses correspond to the GPS stations. The same colors of circle and crosses correspond to the same event.
}
\label{fig:1} 
\end{figure}

For analysis of seismic variations we use the data from “Talaya” seismic station (TLY, 51.681N,103.644E, \cite{II_network,SY_network}). 
The station is located near the point of ionospheric sounding, and both of them are marked at Fig.\ref{fig:1}A by single vertical cross. 
For analysis of the results presented in other publications,
we used the data from seismic stations ARU (56.429N, 58.561E, \cite{II_network}), NKC (50.233N, 12.448E,\cite{SY_network}) and TATO
(24.973N, 121.497E, \cite{IU_network}), also located near the points of corresponding ionospheric observations.

Strong earthquakes at large distances from the epicenter 
sometimes produce in the ionosphere vertical midscale irregularities (multicusp) \cite{Maruyama_2011}.
Their generation is associated with the passage of supersonic surface (Rayleigh) seismic waves below the point
of ionospheric sounding \cite{Rolland_2011,Maruyama_2011,Liu_2011,Chum_2012,Kakinami_2013,Maruyama_2014,Berngardt_2015,Maruyama_2016a,Maruyama_2016b,Liu_2016}, 
and with propagation of resulting acoustic shock wave in the atmosphere and ionosphere.

The amplitude of the effect is determined by the amplitude of the
source seismic vibrations. So selecting the earthquakes that produce the effect
over their magnitude, the distance to them, or other parameters of the earthquake related to epicenter 
is not optimal.
To select the earthquakes participating in the study, we analyzed the
amplitude of the seismic vibrations near the point of ionospheric
observations. 
We estimated the equivalent class $K_{A}$ of the earthquake seismic vibrations by
representing the observed vertical seismic amplitude in logarithmic
scale. This gives us a rough estimation of the amplitude of the source of
the shock wave in the ionosphere at large distances from the epicenter. The
$K_{A}$ is:

\[
K_{A}=10\cdot max\left\{ log_{10}(\widetilde{h})-5\right\} 
\]
where $\widetilde{h}$ is the amplitude of the vertical seismic oscillations
(in nanometers), calculated as the median over
4 seconds to remove noise. 
Table\ref{tab:1} shows the list of detected at TLY station large seismic disturbances with $K_{A}>6$, 
that were analyzed in the paper, and corresponding earthquakes.

Fig. \ref{fig:1}A shows the geometry of investigated earthquakes, 
marked by circles with radius proportional to their $K_{A}$, measured at TLY station. As one can see, 
the earthquakes at larger distances appear with a smaller $K_{A}$, than the earthquakes at smaller 
distances from seismic station. This illustrates the local character of the $K_{A}$ class, related 
with not only the earthquake magnitude, but with the relative position of the epicenter and seismic station also.

\subsection{GPS receivers network}

Investigation of the ionospheric total electron content variations (TEC) in the
vicinity of the earthquake epicenters was made on the basis of phase measurements 
at dual frequency GPS receivers. For every given earthquake epicenter the data from nearby GPS-stations
from International hemodynamic network IGS (http://sopac.ucsd.edu) were used for this study.

TEC variations were calculated by the standard technique used to study the earthquake effects 
\cite{Calais_1994,Afraimovich_2001,Astafyeva_2009}.
In accordance with generally accepted standards, we relate the measured TEC
with the ionospheric point, i.e. with the point of intersection of the ray
"satellite-receiver" with equivalent ionospheric layer at the height of maximal electron density $h_{max}$ 
 \cite{Calais_1994,Afraimovich_2001,Astafyeva_2009,Astafyeva_2013b,Perevalova_2014,Perevalova_2015}. 
In our research we used typical $h_{max}=300km$.  We filtered the initial TEC series to get the variations with periods ranging from 2 to 10 minutes, 
typically intensified by earthquakes in the vicinity of the epicenter \cite{Calais_1994,Afraimovich_2001,Astafyeva_2009,Astafyeva_2013b, Perevalova_2014}
and related with the midscale ionospheric disturbances caused by earthquakes. 

To make more accurate comparison of the intensity of ionospheric disturbances at different stations we transform the inclined TEC variations
into equivalent vertical TEC variations \cite{Afraimovich_2001,Astafyeva_2013b,Perevalova_2015}. 
To differ earthquake effects from regular one we compared the TEC variations at the day of the earthquake with TEC variations
in previous and subsequent days, following to traditional approach \cite{Afraimovich_2001,Perevalova_2014,Perevalova_2015}.

\section{Observations of vertical midscale ionospheric irregularities related with surface seismic waves far from epicenter}

\subsection{F-layer multicusp: ionogrames at the moment of disturbance}

One of the effects observed after a powerful earthquake is 
'multicusp' in the ionosphere, associated with the passage of powerful
shock wave (Mach cone) from supersonic surface seismic waves (Rayleigh waves). 
The propagation of shock wave in the neutral atmosphere generates similar irregularities in the ionosphere through the neutral-ion collisions (see, for example \cite{Maruyama_2014}).
These ionospheric irregularities can be detected by vertical ionosondes, in a form of very specific, short-lived disturbances at the ionograms.
This phenomenon was clearly observed after the Tohoku
earthquake 11/03/2011 \cite{Maruyama_2014,Maruyama_2011,Matsumura_2011,Liu_2013,Berngardt_2015}
and after Chile earthquake 27/02/2010 \cite{Maruyama_2016b}.
Usually this effect is very fast: its total duration is associated with the passage
of the surface seismic waves, and usually do not exceed several minutes.
Therefore, for its diagnosis one needs the ionospheric instruments with
temporal resolution better than 1 minute - a network of GPS-receivers
or fast ionosondes. The weakness of the effect even after strong earthquakes 
makes possible their observation with GPS in either case of the most powerful earthquakes (when disturbances in
the total electron content are significant), or with large spatial
networks of GPS-receivers using special
processing and accumulation techniques presented, for example,  in \cite{Afraimovich_2000,Afraimovich_2001b}.

In comparison with GPS-receiver, covering huge spatial area and allowing us to investigate 
mostly horizontal structure of the irregularities, the ionosonde usually investigates ionosphere 
above its location. Ionosondes are more sensitive instruments for investigating the midscale 
vertical irregularities, and, thus, they are more useful 
than GPS for analysis of vertical ionospheric disturbances after weaker earthquakes or effects at 
larger distances from the epicenter.

The only drawback of ionosondes is their poor temporal resolution of the order
of 10-15 minutes intended mainly for the diagnosis of secondary ionospheric
parameters - the behavior of critical frequencies and the height of
the maximum. Ionosondes with a higher temporal resolution 
were suggested and used for case studies a 
long time ago \cite{Munro_1956,Ponyatov_1999}, but for regular studies 
they appeared quite recently \cite{Reinisch_2009,Akchurin_2011,Harris_2012,Berngardt_2015,Harris_2016,Maruyama_2016a,Maruyama_2016b,SADANKALA}.
IFMCI has the
necessary temporal resolution \cite{Berngardt_2015} and works in
1 minute mode constantly - since 2012, and selectively - since 2011. 
This allowed us to accumulate a huge statistics for monitoring of
ionospheric perturbations associated with earthquakes in this period.

\begin{figure}
\includegraphics[scale=0.48]{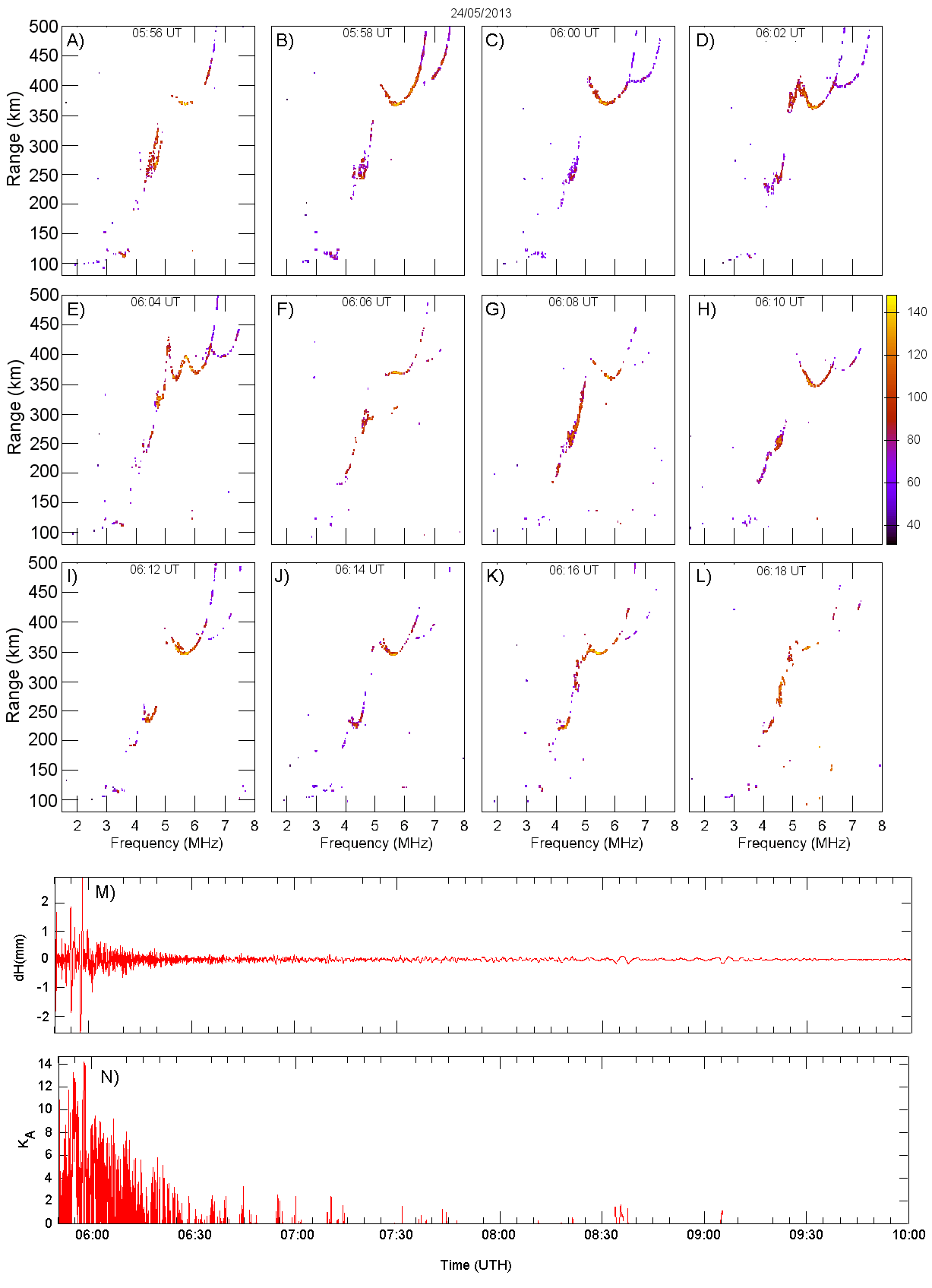}
\caption{
An example of the ionospheric dynamics related to the passage
of the seismic wave at ionograms from IFMCI (after
Okhotsk Sea earthquake, 24/05/2013).
A-L) - consequent ionogrames shown every 2 minutes; 
M) - TLY seismic vertical variations; 
N) - $K_A$ variations.
}
\label{fig:2} 
\end{figure}

Fig.\ref{fig:2} shows an example of the dynamics of vertical ionospheric disturbances ('multicusp') 
at Irkutsk ionosonde data during deep Okhotsk Sea earthquake (24/05/2013), rarely discussed in the literature. 
The most significant multicusp effect is observed up to 300km effective height during 06:02-06:06UT. As one can see, the effect duration is about several minutes. 
This makes it difficult for detection by standard 15-minute ionosondes. The effect is observed about 7-10 minutes 
after the moment of maximal seismic vertical variation. This delay is related with propagation of acoustic signal from the ground to these heights, 
and explains well experimental observations \cite{Maruyama_2011,Matsumura_2011,Liu_2013,Berngardt_2015,Maruyama_2016a,Maruyama_2016b}.

\begin{figure}
\includegraphics[scale=0.30]{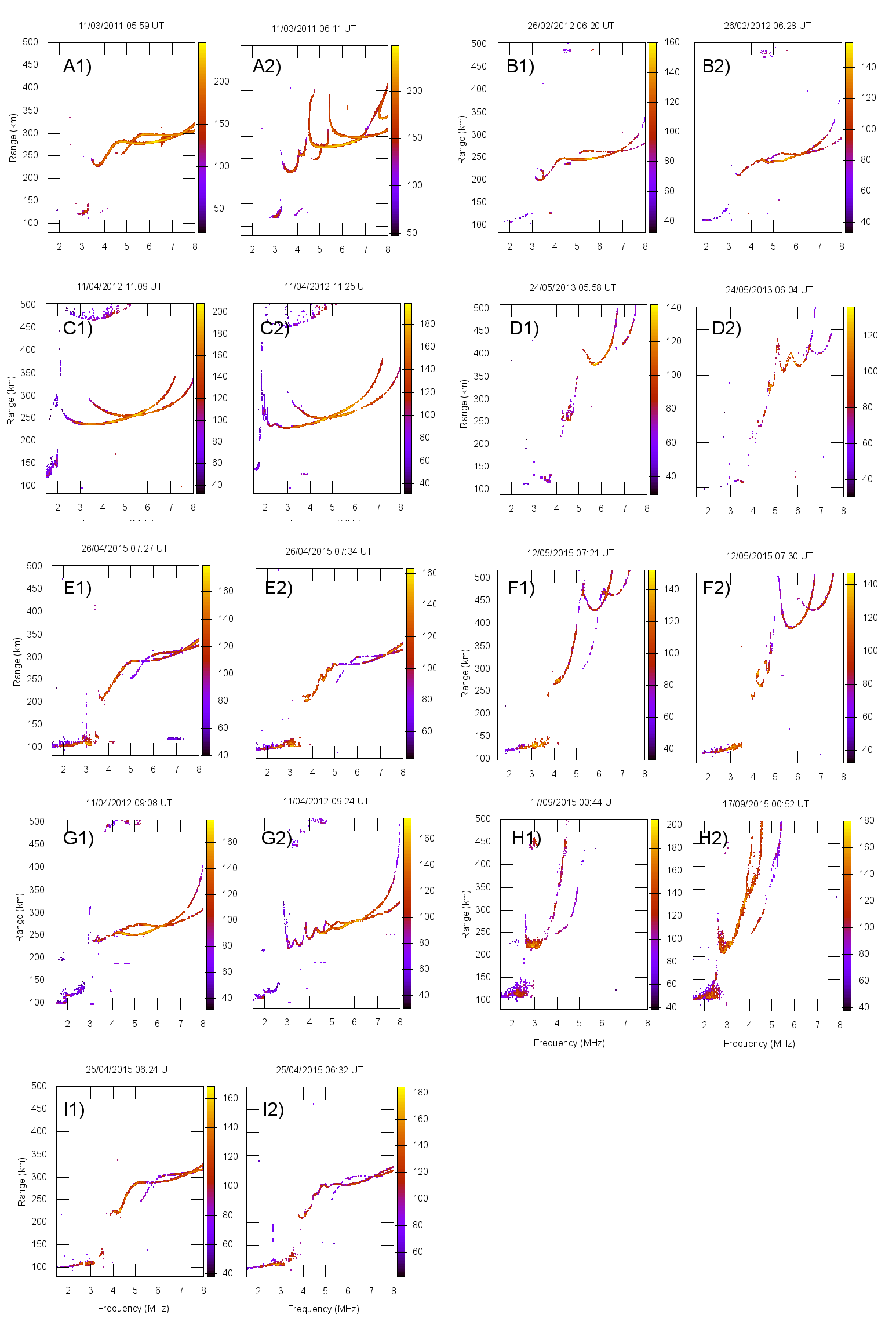}
\caption{
Multicusp ionograms during earthquakes 
A1,2) 11/03/2011 (Mw9.0 Tohoku, Japan), 
B1,2) 26/02/2012 (Mw6.6 Southwestern Siberia, Russia ), 
C1,2) 11/04/2012 (Mw8.4 Northen Sumatra), 
D1,2) 24/05/2013 (Mw8.3 Okhotsk Sea ), 
E1,2) 26/04/2015 (Mw6.7 Nepal ), 
F1,2) 12/05/2015 (Mw7.3 Nepal )
G1,2) 11/04/2012 (Mw8.0 Northen Sumatra)
H1,2) 17/09/2015 (Mw8.3 Coquimbo, Chile )
I1,2) 25/04/2015 (Mw7.8 Nepal )
. Quiet and disturbed profiles marked by 1 and 2 correspondingly. 
}
\label{fig:3} 
\end{figure}

To analyze seismic effects in the ionosphere we analyzed the ionogrames in the period $\pm15 minutes$ from the beginning(end) of most powerful seismic disturbances 
during the investigated period. The ionograms related to each seismic disturbance are shown in Support Materials. As a result, we found 
that the 'multicusp' effect of earthquakes at IFMCI data was observed during the following earthquakes:
11/03/2011 (Mw9.0 Tohoku, Japan), 26/02/2012 (Mw6.6 Southwestern Siberia, Russia), 
11/04/2012 (Mw8.4 Northen Sumatra), 11/04/2012 (Mw8.0 Northen Sumatra), 24/05/2013 (Mw8.3 Okhotsk Sea), 
26/04/2015 (Mw6.7 Nepal), 12/05/2015 (Mw7.3 Nepal), 17/09/2015 (Mw8.3 Coquimbo, Chile), 25/04/2015 (Mw7.8 Nepal). 
Fig.\ref{fig:3} shows the most revealing ionograms during these earthquakes.

From Fig.\ref{fig:3} one can see that sometimes the 'multicusp' effect
is also accompanied by distortion and bifurcation of the F-track (Fig.\ref{fig:3}H2), 
that can be interpreted as horizontal irregularities.
More detailed ionogram dynamics for each earthquake response can also be found in the Support Information.

As it is shown by \cite{Berngardt_2015}, even for a powerful Tohoku 11/03/2011
earthquake the specific periods of perturbation and their
amplitude can be evaluated under assumption of the monotony of the
perturbed electron density profile. Therefore, to evaluate the perturbation amplitude
we processed the ionograms by standard POLAN program (available at \cite{POLAN_Program}), which is  
 traditionally used for multicusp analysis \cite{Maruyama_2011,Liu_2013,Maruyama_2016a,Maruyama_2016b}. 

For automatic processing of the data we developed a simple program set used to convert ionogrames into profiles of plasma frequency. 
The automatic processing was easy to be done due to high quality of Irkutsk ionogrames, already filtered out from neighboring 
public radiostations by original and very powerful technique \cite{Podlesnyi_2014}.
The algorithm for obtaining electron density profiles from ionograms consists of 3 stages.
At the 1st stage, the ionogram is filtered out from rare dot-like noise of different nature.
At the 2nd stage, the track is made over the data by dividing the frequency interval of significant 
signals into 50 subintervals, that is necessary for stable work of POLAN program.
In each subinterval we find the median point over the height and frequency.
At the 3rd stage the median points, collected over subintervals forms the 50-points track, are used as a source 
for POLAN calculations of plasma frequency profile.

\begin{figure}
\includegraphics[scale=0.30]{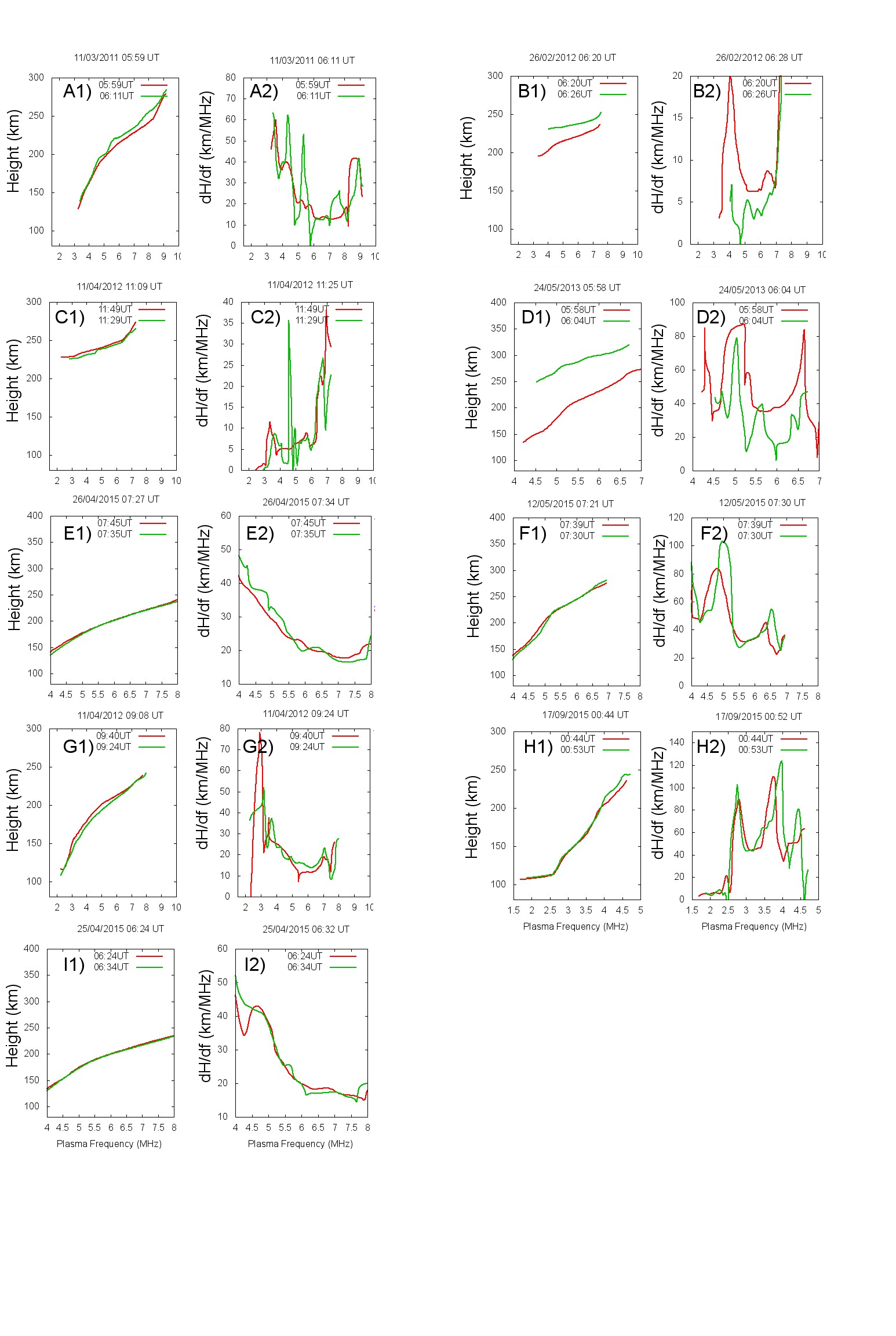}
\caption{
Multicusp profiles during earthquakes 
A1,2) 11/03/2011 (Mw9.0 Tohoku, Japan), 
B1,2) 26/02/2012 (Mw6.6 Southwestern Siberia, Russia ), 
C1,2) 11/04/2012 (Mw8.4 Northen Sumatra), 
D1,2) 24/05/2013 (Mw8.3 Okhotsk Sea ), 
E1,2) 26/04/2015 (Mw6.7 Nepal ), 
F1,2) 12/05/2015 (Mw7.3 Nepal )
G1,2) 11/04/2012 (Mw8.0 Northen Sumatra)
H1,2) 17/09/2015 (Mw8.3 Coquimbo, Chile )
I1,2) 25/04/2015 (Mw7.8 Nepal )
. Quiet and disturbed profiles marked by red and green correspondingly. 
1 and 2 marks height(1) and height derivative(2) correspondingly. 
}
\label{fig:3b} 
\end{figure}

Some results of this automatic processing are shown in Fig.\ref{fig:3b}.
In the figure the plasma frequency profiles are shown, and their differential over the frequency.
From Fig.\ref{fig:3b} one can see that the main perturbations
associated with multicusp can be observed from 140 km up to F2 maximum (for example, Fig.\ref{fig:3b}A), so for
their observations the ionosonde must operate in the appropriate mode, covering reflection heights from 100-140km to F2 maximum.
As one can see, the surface seismic waves lead to formation of mid-scale short-lived plasma frequency variations that disturb initial, relatively smooth ionospheric profile.
The amplitude of these effects is relatively small, and more evident at ionograms (that non-lineary depends on the plasma frequency profile), than in the plasma profile itself.
\subsection{Sporadic E-layer bifurcation as possible effect of surface seismic waves}

\begin{figure}
\includegraphics[scale=0.5]{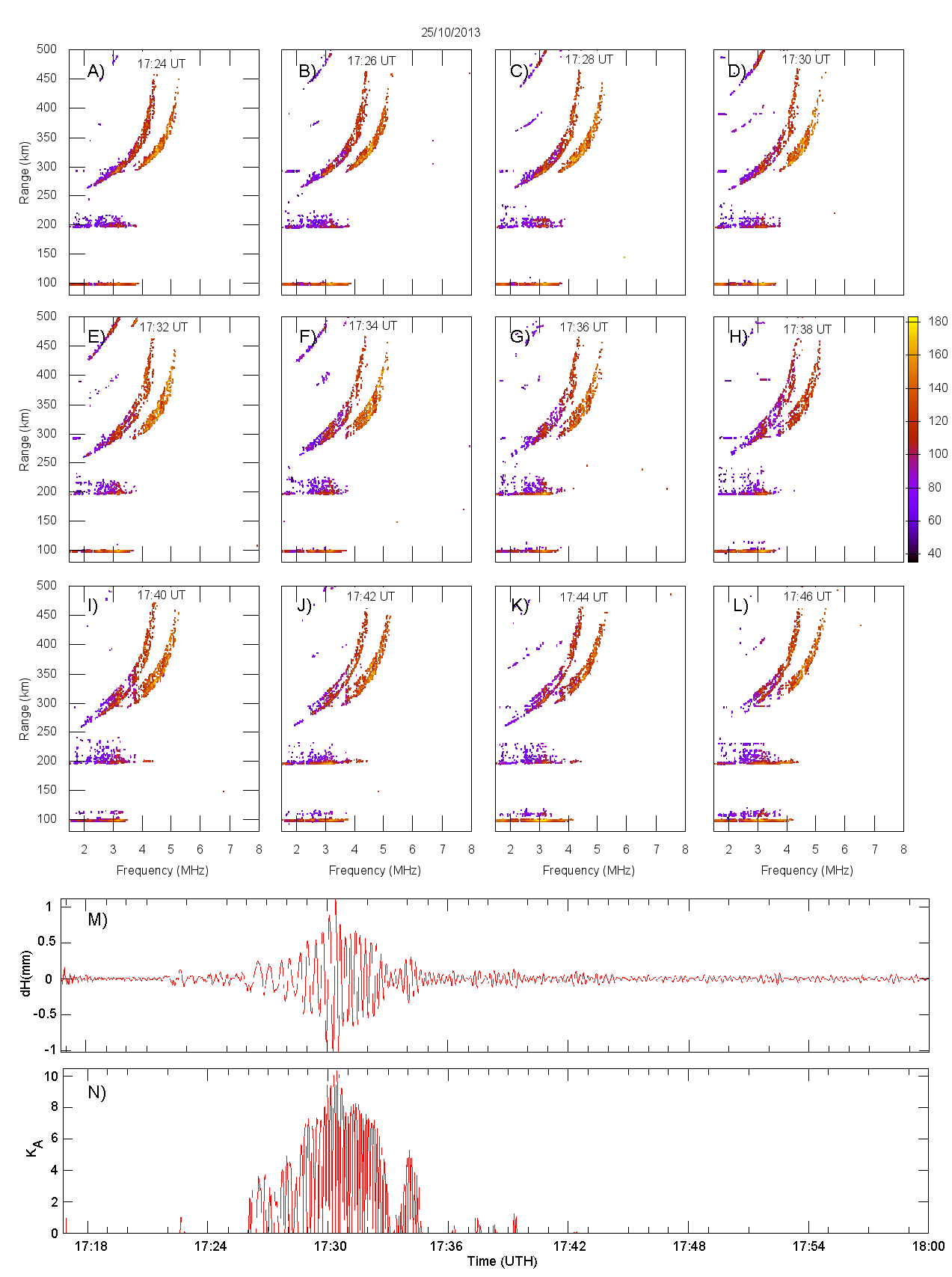}
\caption{
Bifurcation effect in E-sporadic layer after Honshu earthquake 25/10/2013.
A-L) - consequent ionogrames shown every 2 minutes; 
M) - TLY seismic vertical variations; 
N) - $K_A$ variations.
}
\label{fig:4} 
\end{figure}

Another effect, that accompanied some of the earthquakes, was the change
of sporadic E-layer structure, manifested
in the form of bifurcation after the passage of the surface wave.
Fig.\ref{fig:4} shows an example of this effect for the earthquake 07/12/2015
(Mw7.2 Tajikistan).

\begin{figure}
\includegraphics[scale=0.35]{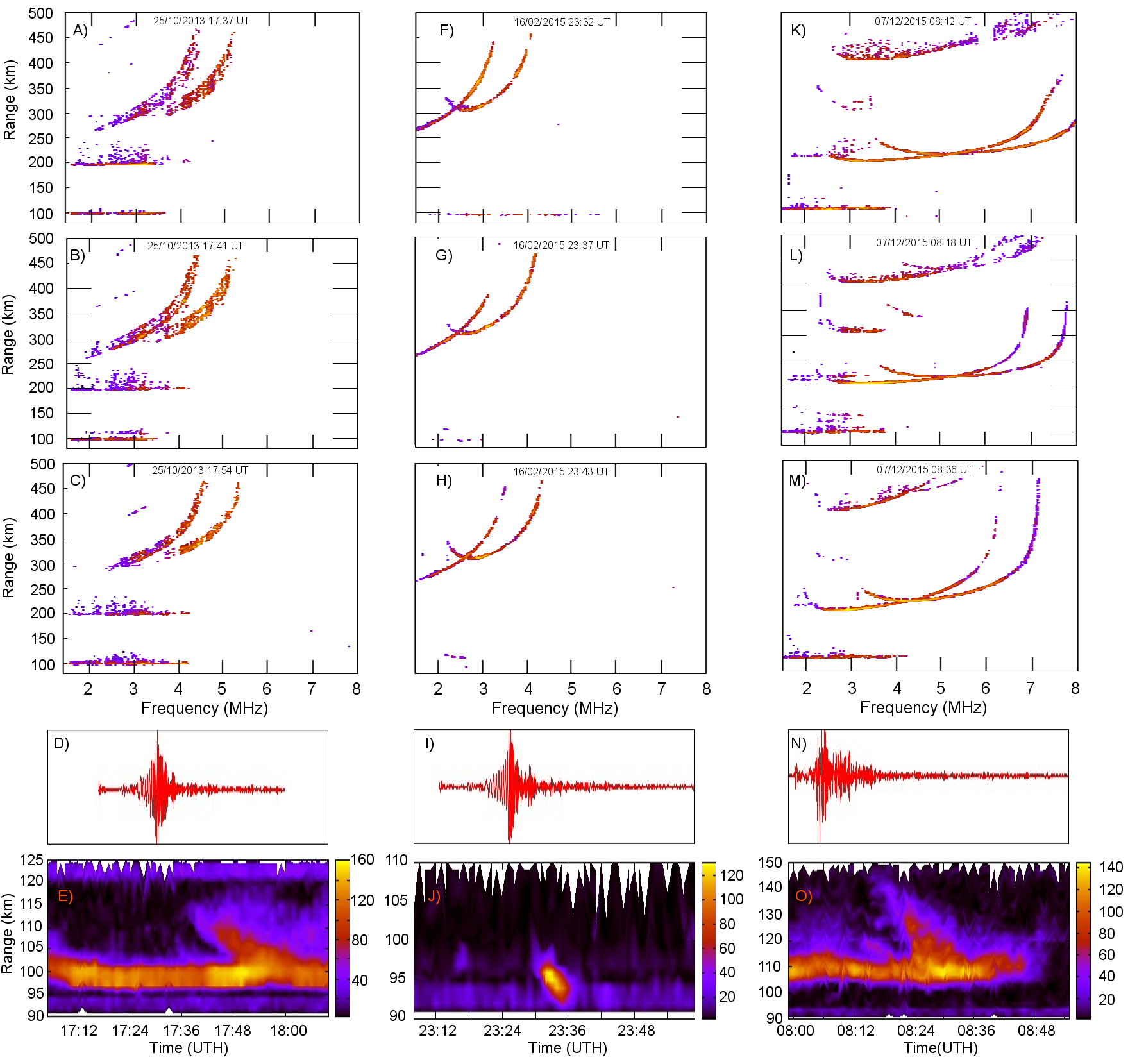}
\caption{
The effect of sporadic E bifurcation
observed in experiments, 25/10/2013 (A-E), 16/02/2015(F-J), 07/12/2015(K-O). 
A,B,C, F,G,H, K,L,M  - ionogrames;
D,I,N - TLY vertical seismic variations;
E,J,O - Intensity of chirp signal reflected from different heights, as a function of the time.
}
\label{fig:5} 
\end{figure}

Fig.\ref{fig:5}A-C,E-G,I-K show the most characteristic ionograms for the earthquakes 
25/10/2013 (Mw7.1 off east coast of Honshu, Japan ), 
16/02/2015 (Mw6.8 near east coast of Honshu, Japan
), 07/12/2015 (Mw7.2 Tajikistan ).

To emphasize this effect, we integrated amplitude at ionograms over frequency for
each given effective height. This technique is close to the height-time-intensity(HTI) technique
\cite{Haldoupis_2006}. Dependence of the resulting amplitude on the effective
height for different earthquakes (when we observe this effect) is
shown in Fig.\ref{fig:5}D,H,L. The figure shows that the main effect is the appearance
of an additional E-sporadic layer overlying the usual E-sporadic layer on
the ionogram. It arises approximately 10-15 minutes after the passage
of seismic disturbance at the point of observation and lasts about 20 minutes. 
After arising, the secondary reflection gradually decreases down to the height of the main E-sporadic track.
At the same time with a decrease of the reflected signal height its
amplitude is increased. 

\section{GPS-observations of IGWs near the earthquake epicenter}

\begin{figure}
\includegraphics[scale=0.6]{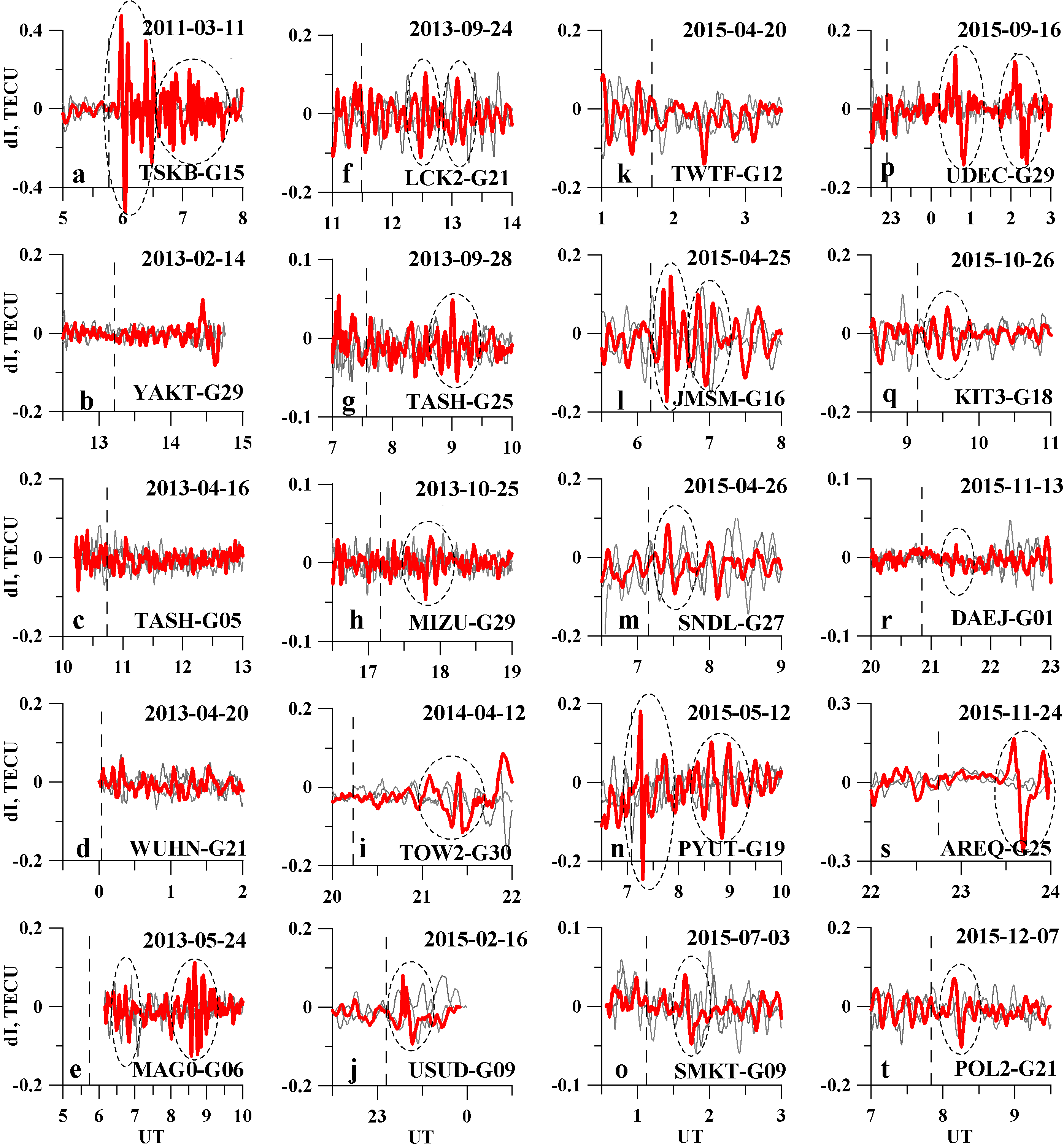}
\caption{
Examples of GPS-TEC disturbances in the vicinity of earthquake
epicenters. The red lines show the variation of the TEC in the day
of the earthquake, gray - the previous and subsequent days. The vertical
dashed line marks the moment of the earthquake. The dashed ellipses
mark disturbances caused by earthquakes.
}
\label{fig:6} 
\end{figure}

Examples of TEC variations in the vicinity of the epicenters
of all considered earthquakes (Table \ref{tab:2}) are shown in Fig.\ref{fig:6}. We
selected the pairs "satellite-receiver" with rays closest to the earthquake's epicenter 
and with larger TEC effect (to take into account aspect dependence of TEC effects).
The TEC variations at those rays are shown in Fig.\ref{fig:6}. 

To estimate the efficiency of each earthquake for generating 2-10min TEC variations, 
we made a preliminary analysis of all the GPS data at the nearest GPS-stations.
The results of the analysis are summarized in Table \ref{tab:2}. The table also
shows the local solar time of the main shock in the epicenter of the
earthquake ("LST"), the focal mechanism of the earthquake ("Fault type") and background
geomagnetic conditions on the day of the earthquake ("M.disturbance"). 

The information about focal mechanisms of the earthquake is obtained from the 
The Global CMT Project(http://www.globalcmt.org).
The geomagnetic conditions are estimated from Dst and AE indexes, obtained from the International Data Centre in Kyoto
(http://wdc.kugi.kyoto-u.ac.jp/wdc). 
In the table we mark the days with disturbed geomagnetic conditions ("disturbed")
and with magnetic storms ($Dst<-50nT$, "storm"), based on the behavior of these indexes.

From the Table \ref{tab:2} one can see that during most part of the earthquakes the 
geomagnetic conditions were quiet. So ionospheric irregularities related with geomagnetic disturbances 
has not prevent us to detect TEC-GPS response to earthquakes.

We classify TEC disturbances generated by the earthquakes into 3 following categories.

"Strong response" has relatively high amplitude ($\ge 0.2$ TECU) (Fig.\ref{fig:6}a,l,m,n,p)
and recorded at many rays "receiver-satellite".
Such responses are registered after five earthquakes: 11/03/2011, 25/04/2015, 26/04/2015, 12/05/2015, 16/09/2015. 
The magnitude of these earthquakes varies from 6.7 to 9.0, and all of them had reverse fault
focal mechanism. It is important to note about very successful geometry
of the measurements during all of these earthquakes: the most part of the "receiver-satellite"
rays were close to or directly above the epicenter. 

In this category the most powerful earthquake 11/03/2011 (Tohoku, Japan) should be emphasized.
It causes the most intense and prolonged ionospheric disturbances.
Ionospheric effects of this earthquake are investigated in a huge amount of papers (see for example \cite{Astafyeva_2011,Liu_2011, Nishitani_2011}).
Recently, more attention is paid to modeling of these effects \cite{Kherani_2012,Kakinami_2013,Shinagawa_2013}
and to the study of the propagation of the disturbances at large distances
\cite{Jin_2014,Berngardt_2015,Tang_2015}.

"Weak response" is the response that has relatively low amplitude (0.1-0.15
TECU) and recorded at several (2 to 5) rays 'satellite-receiver'. 
The shape and amplitude of such disturbances is high enough
to detect it from the background TEC variations level. 

In this category the deep earthquake 24/05/2013(Okhotsk Sea earthquake, Fig.\ref{fig:6}e) 
should be emphasized. 
It was very  powerful earthquake (Mw8.3) with a predominance of vertical displacements 
in the epicenter (normal fault). It attracted the attention of seismologists,
because felt at unusually large (10,000 km) distance from the epicenter
\cite{Ye_2013,Zhan_2014}. 
The Earth's surface displacement in the Sea of Okhotsk area caused by this earthquake
were also studied using GPS-data \cite{Steblov_2014}. 
However, due to the very large depth of the epicenter, the impact of the Okhotsk
Sea earthquake to the ionosphere looks very weak. Perhaps that is why
the ionospheric effects of the earthquake were not discussed too much
in the literature. 
We could not find any work devoted to the GPS-TEC or ionospheric perturbations
caused by this event on nearby region. 
During the present study we detected some TEC perturbations,
most likely associated with the Okhotsk Sea earthquake, but only at two rays
(MAG0-G06, MAG0-G16), closest to the epicenter of the earthquake.

"No response" category in the Table \ref{tab:2} means that we
were unable to identify any significant TEC disturbances related to
the earthquake at any of the rays "receiver-satellite"
(Fig.\ref{fig:6} b, c). 
The effects were not observed after two earthquakes: 14/02/2013 and 16/04/2013. 
In our opinion the disturbance observed in 14/02/2013 (Fig.\ref{fig:6}b) is not
a response to the earthquake, but should be associated with the intersection
by the ray "receiver-satellite" by the sharp
electron density gradient, perhaps the highlatitude trough. 
The analysis of the original (unfiltered) series of TEC variations leads us to this conclusion. 
As the analysis has shown, the series have a bend ("hook"), which is a characteristic for the cases when the 
beam crosses the steep gradient of the electron density, for example, the terminator or  boundary 
of ionospheric trough.
To "No response" category we also attributed the events when we
detect a weak perturbations at one or two rays  with the shape close to the shape 
associated with earthquake-generated perturbations usually \cite{Calais_1994,Afraimovich_2001,Astafyeva_2009, Astafyeva_2013b,Perevalova_2014,Perevalova_2015} 
(Fig.\ref{fig:6} d, k). However, due to a number of reasons (the absence of such disturbances
on the other rays, a significant propagation velocity exceeding sound speed, etc.) 
there is no assurance that these perturbations can be caused by an earthquake (for example, in case of 20/04/2013 and 20/04/2015). 

No GPS-TEC response to the earthquake may have several causes. 
First of all,  the earthquakes with small magnitude
and with predominance of horizontal displacement in the focus (strike-slip faults) that produce very small effects with amplitudes lower than background effects. 
As it was shown in \cite{Perevalova_2014}, after the earthquakes
with magnitudes $Mw<6.5$ appreciable TEC wave-like disturbances are
not observed, and in the case of strong earthquakes ($Mw>6.5$) responses
are more pronounced after the events with substantial vertical component in the focus (normal or reverse fault). 
Another important cause is the geometry of GPS-measurements. 
In most cases, the lack of response at the beam path "receiver-satellite" 
may be related with a large distance from the ionospheric point to the epicenter
(the mark "rays far away" in Table \ref{tab:2}).

For "strong response" and "weak response" categories we estimated the propagation
velocity of disturbances. The estimate is done for the first
impulse response. The velocity $V$ is calculated from a simple linear
propagation model $V=dS/dt$, where $dS$ is the line-of-sight distance between
the epicenter and the ionospheric point at which the response was
recorded, $dt$ is the delay between the TEC variation maximum moment 
and the moment of the earthquake (shown in Table\ref{tab:2}).
The resulting values of $V$ are summarized in Table \ref{tab:2}. 
The calculated propagation velocity varies between 150 and 600 m/s. 
This indicates that these perturbations can be associated
with propagation of internal atmospheric waves \cite{Shinagawa_2007,Occhipinti_2008,Matsumura_2011,Kherani_2012}.

~

\section{Discussion}

\subsection{Class of Reyleigh wave acoustical efficiency $K_{W}$}

\begin{figure}
\includegraphics[scale=0.33]{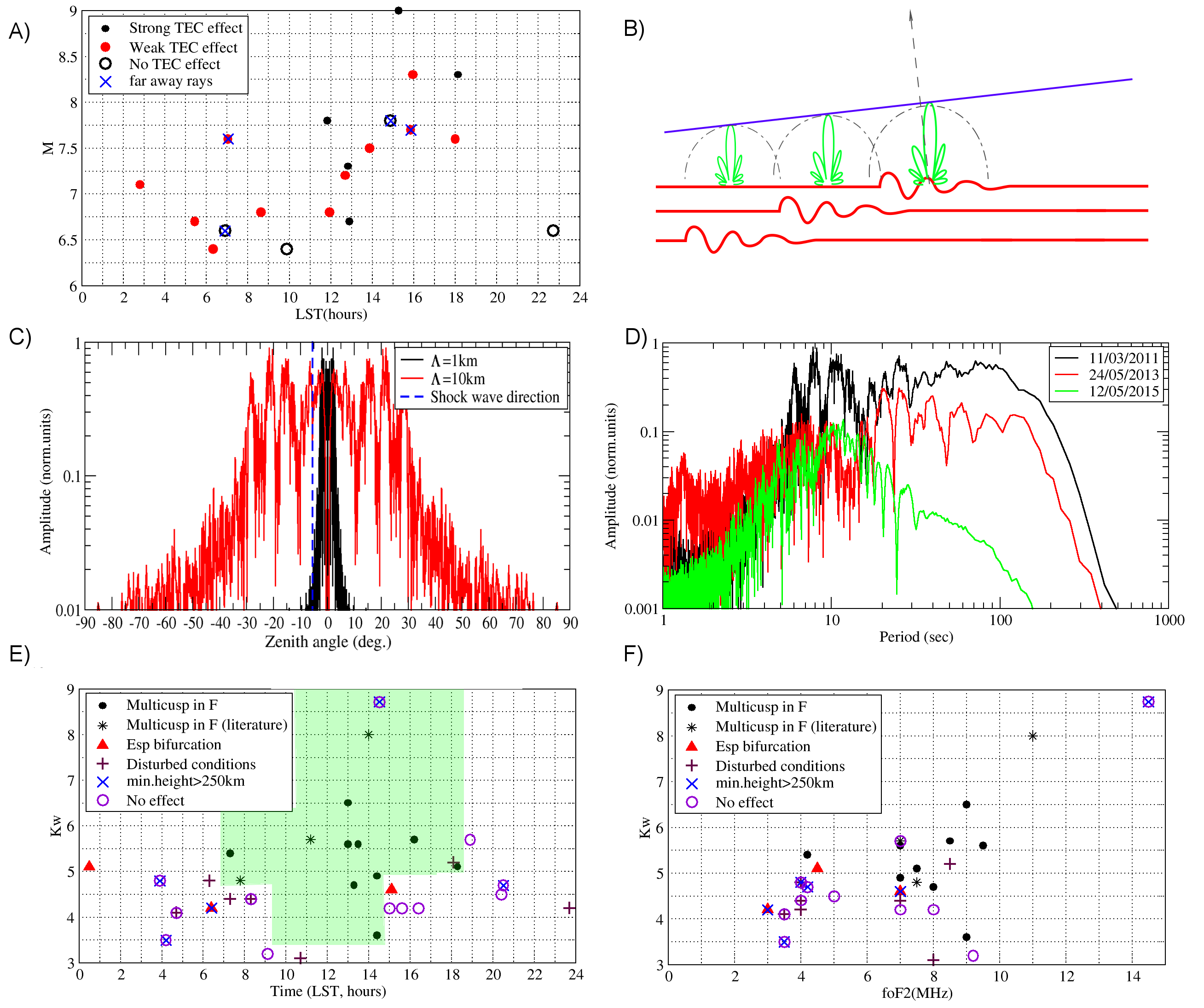}
\caption{
A) The ionospheric effects of earthquakes in the vicinity
of epicenter detected by TEC-GPS technique, as function on the earthquake magnitude and local solar time; B) Scheme of shockwave (Mach cone) formation with taking into account acoustic radiation pattern.
C) Intensity of acoustic wave as a function of zenith angle (acoustic radiation pattern). Calculated from TLY vertical seismic oscillations for 11/03/2011 (Tohoku) earthquake.
Red line corresponds to 10 km wavelength, black line - to 1 km wavelength. Blue dashed line corresponds to the zenith angle of shockwave propagation.
D) The spectrum of seismic oscillations for earthquakes producing multicusp
in Irkutsk. E) Types of observed ionospheric effects related with surface seismic waves in 2011-2016 as a function of local time and acoustic efficiency class $K_W$. Green area marks the area with all the multicusps observed.
F) Types of ionospheric responses as a function of foF2 and acoustic efficiency class $K_W$.
}
\label{fig:7} 
\end{figure}

As the analysis of GPS-TEC measurements shows above, near the epicenter 
the most part of earthquakes were accompanied by TEC-variations (see Fig.\ref{fig:7}A). From the other side, 
the multicusp effect at large distances accompanies only several earthquakes.

So the question naturally arises, how can we estimate the effectiveness
of a given surface seismic wave for the formation of those or other effects
in the ionosphere. To estimate this, we introduce the class
of Rayleigh wave acoustical efficiency (RWAEC) that is determined by the
amplitude of the local seismic vibrations below the point of ionospheric
observations. 

It can be shown that for distributed acoustic wave source the Mach cone
amplitude $A_{acoustic}$ can be estimated by multiplying Mach cone amplitude from moving point
source to the acoustic radiation pattern of distributed source: 
\begin{equation}
A_{acoustic} \approx G_{Mach} \cdot g(asin(\frac{C_s}{V_{Rayleigh}}))
\end{equation}
where 
 $G_{Mach}$ is the amplitude of acoustic signal in the shock wave, generated by dot-like (isotropic) acoustic source moving with supersonic speed; 
 $g(\alpha)$ is the acoustic radiation pattern of the distributed source itself as a function of zenith angle $\alpha$, 
 calculated from relation between sound speed $C_s$ and supersonic source speed $V_{Rayleigh}$ (supposed to be 3.5km/s).

The radiation pattern $g(\alpha)$ is defined by the
spatial structure of the propagating seismic wave.  This approximation does not take into account the phase radiation pattern. The exact
spatial wave structure is unknown for us, but we can assume that the wave has some spatial shape,
that moves without dispersion with the supersonic speed of seismic wave.

Based on this approximation we can use temporal variations of the seismic signal at a given station
to estimate the spatial shape of an equivalent acoustic source.
In this case we can calculate the radiation pattern associated with the spatial distribution
of seismic oscillations, and estimate the amplification of the Mach cone as a function of seismic signal shape. 
It should be noted that since the seismic wave velocity is much higher than acoustic sound speed in the atmosphere, 
one needs to take into account the radiation pattern only at the angles close to the perpendicular to the Earth's surface 
(i.e. perpendicular to the radiation plane).

The principle of the formation of a Mach cone with taking into account
the acoustic signal radiation pattern is illustrated in Fig.\ref{fig:7}B.

The intensity of the acoustic signal in the far zone of the acoustic antenna (that is approximately valid for this case)
is defined by radiation pattern. The radiation pattern $g(\overrightarrow{k})$ of distributed sound source 
in its far field zone is defined \cite{Smaryshev_1973} as:

\begin{equation}
g(\overrightarrow{k})=\int\Omega(\overrightarrow{r})e^{i\overrightarrow{k}\overrightarrow{r}}d\overrightarrow{r}
\end{equation}

where $\Omega(\overrightarrow{r})$
is vibrational velocity of the sound source (we suppose the movements
to be strictly vertical). In one-dimensional, steady-state case, we
can estimate the vertical velocity of the surface oscillations from
the experimental data as

\begin{equation}
\Omega(r)=\left\{ \frac{\widetilde{h}(t)}{dt}\right\} _{t=(r-r_{0})/V_{seismo}}
\label{eq:spectrum}
\end{equation}

In this case, the wave vector of the radiation $\overrightarrow{k}$ 
is close to the perpendicular to the plane of the source $\overrightarrow{r}$
and their scalar product is small. In the one-dimensional case, for plain Earth approximation, along
the direction of the Rayleigh wave motion (this case corresponds to
a significant distance between observation point and the epicenter
of the earthquake, when the seismic wave can be considered having
not spherical but plane front), the acoustic radiation pattern
can be estimated from one-dimensional spectrum:

\begin{equation}
g(\overrightarrow{k})=i k_{eff}V_{seismo}\int\widetilde{h}(r)e^{ik_{eff}r}dr
\end{equation}

where $k_{eff}=\overrightarrow{k}\frac{\overrightarrow{r}}{r}$ is a wavenumber projection to the acoustic source plane.

Fig.\ref{fig:7}C shows examples of radiation pattern, calculated for Tohoku (11/03/2013) earthquake for two 
wavelengths - $\Lambda=1km$ (black line) and $\Lambda=10km$ (red line), as a function of zenith angle. The dashed line marks 
the direction of shockwave propagation. As one can see, the antenna pattern 
is different for different wavelengths. So we can find the wavelength, for which the amplitude of acoustical signal becomes maximal in shockwave direction.
The amplitude corresponds to the maximum of the spectrum (\ref{eq:spectrum}). So searching the
wavelength most effective  for generating acoustical signal in the first approximation corresponds to the search of 
maximum in the spectrum (\ref{eq:spectrum}).

Thus, for fixed height of the expected effect $R$ and excluding the
effects of the sound propagation in the neutral atmosphere and the
energy transformation from the neutral component to the charged one,
the spectral energy class of the acoustic signal caused by seismic
vibrations can be estimated from the maximal amplitude of radiation
pattern, and consequently
from the spectrum of the derivative of seismic vibrations like:

\begin{equation}
K_{w}=log_{10}\left\{ max\left\{ \left|\frac{\omega}{2\pi}\int_{T}\widetilde{h}(t)e^{-i\omega t}dt\right|^{2}\right\} \right\} -10
\label{eq:kw}
\end{equation}

where $\widetilde{h}$ in nanometers. For calculating the spectrum in (\ref{eq:kw}) we used 8192 and 16384 point fast Fourier transform.
Fig.\ref{fig:7}D shows a comparative analysis of the spectra of the derivative
of the vertical vibrations during several earthquakes (11/03/2011,
24/05/2013 and 12/05/2015) that are close in Irkutsk local solar
time, as well as close in season. One can see that the spectra (and acoustic radiation patterns)
differs significantly, and this can explain the difference in the amplitude of the effect,
observed in the ionosphere.

~

\subsection{Statistical dependences}

~

The Fig.\ref{fig:7}E and Table \ref{tab:3} summarize the main effects of earthquakes observed
by IFMCI as a function of local solar time and Rayleigh
wave acoustic efficiency class $K_{W}$. One can see from the Fig.\ref{fig:7}E that
at night multicusp effect is not observed. This can be due to the
fact that for nighttime low electron density the standard ionograms
start from 250km altitude. For relatively weak $K_{W}<3.6$
multicusp is not observed even in the daytime. For relatively high $K_{W}\ge4.8$ multicusp is observed nearly regularly at daytime [07:00-17:00]LST.
It should be noted that at the 16:00 and 01:00 LST multicusp is not observed,
although tracks observed at ionograms allow us to detect such effect.
We can make a conclusion that to observe multicusp in the F-layer
is too difficult at night. This can be explained qualitatively by the need
of more energy for seismic oscillations at nighttime to generate waves
in high F-layer than at daytime to generate the effects in the lower
F1 or E-layer. This is due to the fact that lower electron density produces less neutral-electron
collisions, that are the main agents for energy transfer from neutral
acoustic wave to electron density variations \cite{Chum_2012,Lastovicka_2006,Maruyama_2014}. 

To verify our conclusions about the power and daily features of multicusp
observations we also analyzed the results from some of the most detailed
papers. In particular, we calculated $K_{W}$ index for the observations of ionospheric multicusp reported in
(\cite{Chum_2016,Maruyama_2016a,Liu_2011}). 
The calculation of  $K_{W}$ was made from the seismic
vibrations observed at seismic stations nearby the locations of observed
ionospheric effects (the ARU seismic station (Sverdlovsk region, Russia),
TATO seismic station (Taiwan) and NKC seismic station(Czech Republic)).
The results are also summarized in the Table \ref{tab:3} and Fig.\ref{fig:7}E. 
These cases are marked by asterisks and confirm the obtained 
results. 

In the morning, evening and night time (LST $>$ 15:00,LST $<$ 07:00) we observed 
cases of bifurcation effect in the sporadic E-layer. This track separation can be explained
by the formation of an additional horizontal layer above the regular sporadic-E
that causes the appearance of an additional reflection point. 
One can assume several variants of how to explain the observed bifurcation
effect in the sporadic E-layer. 

We assume that we observe a downward movement
of irregularities due to translucency of sporadic E-layer, by analogy
with \cite{Haldoupis_2006,Haldoupis_2012}. In this case, the downward movement occurs
with nearly wind speeds. 

The presented observations of GPS-TEC disturbances in the vicinity
of earthquake epicenters confirm the data obtained in previous studies
\cite{Perevalova_2014,Astafyeva_2013b}. With the increasing of the earthquake magnitude
, in general, an increase in TEC amplitude perturbations
is observed as well. The TEC response is also affected by the mechanism of
the earthquake source: after the earthquake, in the epicenter of which
a vertical displacements are dominated, the TEC variations are intensified
and are detected at a large number of rays "satellite-receiver".
The geometry of measurements also plays an important role: the absence
of rays "satellite-receiver" near the
epicenter makes it difficult to detect responses even after earthquakes
with magnitudes Mw$>$7. It may also be noted that after strong earthquakes
(Mw$>$8), in addition to the first pulse associated with the main shock,
a number of secondary vibrations is observed. They are caused,
in our opinion, by the generation of eigen oscillations of the atmosphere
with different periods (IGWs) \cite{Lognonne_1998,Artru_2001}. 

The dependence of TEC effects on the local time has not been identified
(see Table\ref{tab:2}). The diurnal variation of effects suggests that generation
of IGWs more likely do not depend on the local time. The dependence
of intensity of IGWs effects on the nature of earthquakes was discussed,
for example, in \cite{Astafyeva_2014}.

This means that the found dependence of multicusp from the local time 
that is not detected in the epicenter vicinity. It significantly depends on the 
specific mechanism of multicusp generation by the shock wave. This can be 
qualitatively explained by the dependence of the ion-neutral collisions 
mechanism on the product of the background neutral density and background 
electron density, that is maximal in a daytime.

\section{Conclusion}

In the paper the statistical analysis of ionospheric effects of earthquakes
that occured in 2011-2016 according to the Irkutsk fast monostatic chirp ionosonde was made.
To control the process of neutral-ionospheric interaction in the vicinity
of epicenter the data from GPS-receivers were also analyzed for each
of these earthquakes.

To estimate the ionospheric efficiency class for seismic disturbances
in the far field of epicenter, that cause propagation of the shock
cone (Mach cone) we proposed the logarithmic index $K_{W}$
(\ref{eq:kw}), based on finding the maximal amplitude of spectral
power fluctuations. From a physical
point of view, the index allows us to estimate the maximal amplitude
of the shock wave (Mach cone) based on spatial distribution of seismic
oscillations and their vertical velocities. So the index depends on
the amplitude of the acoustic effects associated with the
passage of seismic surface wave. The analysis shows that the characteristic
index value, from which IFMCI can see multicusp
effect in the ionosphere (at daytime [7:00-17:00]LST  $K_W\ge4.7$). The bifurcation of sporadic E-layer
can be observed at nighttime at $K_W>3.6$.

It is shown that a multicusp effect according IFMCI 
data has a rather pronounced daily dependence, intensified in local
daytime hours, with absence of characteristic dependence on foF2.
This allows us to suggest that the possibility of observing the effect
is likely due not to the intensity of the primary F2 layer , but due
to other mechanisms at lower heights. It is shown that when 
ionograms start above 250km (in the morning, evening and night) the multicusp
effect in Irkutsk is not observed, this corresponds well with results of \cite{Chum_2016}. 

It is shown that this effect is not associated with daily dependence
of the generation of ionospheric disturbances in the vicinity of the earthquake
epicenter, estimated by GPS data. This suggests that the efficiency
of generation of irregularities on the Mach shock wave and in the vicinity
of earthquake epicenter are apparently different.

On the example of deep earthquake in the Okhotsk Sea (25/10/2013,
depth about 600 km) it can be assumed that the effects
of Rayleigh wave in the case of deep earthquakes may be more noticeable
than IGW effects in the vicinity of earthquake epicenter. Thus, in spite
of the daily dependency effects, the effects from Rayleigh waves are
an additional way to study the ionospheric response to earthquakes,
because they can sometimes produce more strong effects than IGWs generated
in the epicenter.

It is shown that after passing the Rayleigh wave sometimes can be
observed a bifurcation in sporadic E-layer, observed as secondary E-layer
arised and moving downward to basic sporadic E. This can be associated
with formation of vertical irregularities and their dynamics under
the influence of the dynamics of the neutral atmosphere. 

As a result of the work it is shown that fast IFMCI 
is a very sensitive instrument for investigating of rapid ionospheric
effects related to the earthquakes with $K_{W} \leq 3.6$ during the 
day, which roughly corresponds to the characteristic magnitudes of
distant earthquakes above $M\geq6.5$. This makes the ionosonde a
convenient tool for the diagnosis of various processes that occur
during seismo- ionospheric interaction.

The obtained results allow us to suggest $K_W$ and local solar time as 
effective parameters for searching the multicusp effects in the ionosphere 
related with surface seismic waves.


%
%
%
%
%
%
%

\section{Acknowledgments}
We are greateful to ISTP SB RAS stuff: to Dr.Lebedev V.P. and Dr.Tashilin A.V. for fruitful discussion, to Ivanov D.V. and  Salimov B.G. for preparing ionograms for analysis.

We are grateful to Scripps Orbit and Permanent Array Center
(SOPAC) for the provision of data from the global network of GPS receivers.

We are greateful to IRIS/IDA Seismic Network (II),  Global Seismograph Network (GSN - IRIS/USGS) (GSN,IU), SY - Synthetic Seismograms Network (SY) for providing seismic data.

Irkutsk fast monostatic chirp ionosonde data is property of ISTP SB RAS, contact Oleg I.Berngardt (berng@iszf.irk.ru). The functionality of IFMCI was
supported by FSR program \#II.12.2.3.

The work was done under financial support of the project \#0344-2015-0019
"Study of the lithosphere-atmosphere-ionosphere system in extreme
conditions" of the program of Presidium of RAS,  
grant NSH-6894.2016.5 of the President of State Support for Leading
Scientific Schools.

\newpage

\newpage
\begin{table}
\tiny
\begin{tabular}{|c|c|c|c|c|c|}
\hline 
{$K_{A}$} & {Date/time} & {Epicenter} & {Depth} & {M} & {Location}\\
\hline 
\hline 
{17} & {2011-03-11 05:46:23.0} & {38.30N,142.50E} & {22} & {9.0} & {Honshu, Japan (Tohoku)}\\
\hline 
{16} & {2012-02-26 06:17:19.0} & 	{51.72N,95.99E}&{10}&{6.6}&{Southwestern Siberia, Russia} \\
\hline
{12} & {2012-04-11 08:38:35.0(a)} & 	{2.37N,93.17E}&{10}&{8.4}&{Northern Sumatra}\\
\hline
{8} & {2012-04-11 10:43:09.0(b)} & 	{0.81N,92.45E}&{10}&{8.0}&{Northern Sumatra}\\
\hline
{7} & {2012-12-07 08:18:23.0} & 	{37.92N,144.02E}&{30}&{7.3}&{Honshu, Japan}\\
\hline 
{11} & {2013-02-14 13:13:53.0} & {67.62N,142.66E} & {10} & {6.6} & {Sakha, Russia}\\
\hline 
{10} & {2013-04-16 10:44:20.0} & {28.14N,62.08E} & {87} & {7.8} & {Iran-Pakistan border}\\
\hline 
{13} & {2013-04-20 00:02:48.7} & {30.33N,102.99E} & {20} & {6.6} & {Sichuan, China}\\
\hline 
{14} & {2013-05-24 05:44:48.0} & {54.91N,153.34E} & {598} & {8.3} & {Okhotsk Sea}\\
\hline 
{13} & {2013-09-24 11:29:49.0} & {27.07N,65.56E} & {20} & {7.7} & {Pakistan}\\
\hline 
{8} & {2013-09-28 07:34:06.0} & {27.24N,65.56E} & {10} & {6.8} & { Pakistan}\\
\hline 
{10} & {2013-10-25 17:10:17.0} & {37.22N,144.69E} & {10} & {7.1} & {Honshu, Japan}\\
\hline 
{8} & {2014-04-12 20:14:37.0} & {11.26S,162.26E} & {10} & {7.6} & {Solomon Islands}\\
\hline 
{8} & {2015-02-16 23:06:27.4} & {39.87N,142.94E} & {15} & {6.8} & {Honshu, Japan}\\
\hline 
{6} & {2015-04-20 01:42:55.5} & {24.21N,122.52E} & {7} & {6.4} & {Taiwan Region}\\
\hline 
{16} & {2015-04-25 06:11:26.9} & {28.24N,84.74E} & {15} & {7.8} & {Nepal}\\
\hline 
{8} & {2015-04-26 07:09:10.3} & {27.86N,86.08E} & {15} & {6.7} & {Nepal}\\
\hline 
{12} & {2015-05-12 07:05:19.0} & {27.89N,86.17E} & {10} & {7.3} & {Nepal}\\
\hline 
{15} & {2015-09-16 22:54:31.8} & {31.55S,71.58W} & {20} & {8.3} & {Coquimbo, Chile}\\
\hline 
{15} & {2015-09-17 23:18:43.1} & {31.55S,71.68W} & {40} & {7.0} & {Coquimbo, Chile}\\
\hline 
{11} & {2015-10-26 09:09:32.0} & {36.48N,70.91E} & {207} & {7.5} & {Hindu Kush, Afghanistan}\\
\hline 
{8} & {2015-11-13 20:51:35.2} & {30.95N,129.00E} & {10} & {6.7} & {Kyushu, Japan}\\
\hline 
{12} & {2015-12-07 07:50:07.2} & {38.18N,72.91E} & {30} & {7.2} & {Tajikistan}\\
\hline 
{7}	& {2016-01-30   03:25:09.8}	& {54.03N,158.54E}  &	{159}	& {7.2}	 & {Kamchatka, Russia} \\
\hline
{13}	& {2016-03-02   12:49:48.0}	& {4.89S,94.25E}  &	{20}	& {7.8}	 & {Sumatra, Indonesia} \\
\hline 
{11}	& {2016-04-15   16:25:07.3}	& {32.85N,130.57E}  &	{10}	& {7.0}	 & {Kyushu, Japan} \\
\hline
{8}	& {2016-04-16   23:58:37.2}	& {0.40N,79.89W}  &	{20}	& {7.8}	 & {Ecuador} \\
\hline
{10}	& {2016-07-29   21:18:30.2}	& {18.58N,145.51E}  &	{267}	& {7.7}	 & {N. Mariana Islands} \\
\hline 
\end{tabular}
\caption{List of seismic disturbances, participated in the analysis,
and the corresponding earthquake (according to the European-Mediterranean
Seismological Centre http://www.emsc-csem.org).}
\label{tab:1}
\end{table}

\newpage

\newpage

\newpage

\newpage

\newpage

\newpage

\newpage
\begin{table}
\tiny
\begin{tabular}{|c|c|c|c|c|c|c|c|}
\hline 
EQ date & UT & M & LST  & M.disturbance & Fault type & V(m/s)  &  Effect \\
\hline 
\hline 
2011-03-11  & 05:46:23 & 9.0 & 15.267 & AE disturb. & reverse  & 535  & strong response\\
\hline 
2013-02-14  & 13:13:53 & 6.6  & 22.728  & quiet & reverse+strike-slip  & -  & no response\\
\hline 
2013-04-16  & 10:44:20  & 7.8  & 14.872 & quiet & normal+strike-slip  & -  & no response, rays far away\\
\hline 
2013-04-20  & 00:02:48  & 6.6  & 6.899 & quiet  & reverse+strike-slip & -  & no response, rays far away\\
\hline 
2013-05-24  & 05:44:48  & 8.3  & 15.956 & quiet  & normal & 165  & weak response\\
\hline 
2013-09-24  & 11:29:49  & 7.7  & 15.854 & AE disturb. & strike-slip  & 368  & weak response, rays far away\\
\hline 
2013-09-28  & 07:34:06  & 6.8  & 11.938 & quiet & strike-slip  & 316  & weak response\\
\hline 
2013-10-25  & 17:10:17  & 7.1  & 2.813 & quiet  & normal  & 152  & weak response\\
\hline 
2014-04-12  & 20:14:37  & 7.6 & 7.05  & Dst storm & normal  & 304  & weak response, rays far away\\
\hline 
2015-02-16  & 23:06:27   & 6.8 & 8.629  & Dst storm & reverse  & 460  & weak response\\
\hline 
2015-04-20  & 01:42:55  & 6.4  & 9.868  & Dst disturb. & reverse & -  & no response\\
\hline 
2015-04-25  & 06:11:26  & 7.8  & 11.83 & quiet & reverse  & 505  & strong response\\
\hline 
2015-04-26  & 07:09:10  & 6.7 & 12.889 & quiet & reverse  & 394  & strong response\\
\hline 
2015-05-12  & 07:05:19  & 7.3 & 12.828 & AE disturb. & reverse  & 525  & strong response\\
\hline 
2015-07-03  & 01:07:46  & 6.4 & 6.324  & quiet & reverse  & 286  & weak response\\
\hline 
2015-09-16  & 22:54:31  & 8.3 & 18.128 & AE disturb. & reverse  & 145  & strong response\\
\hline 
2015-10-26  & 09:09:32  & 7.5 & 13.877 & quiet & reverse  & 423  & weak response\\
\hline 
2015-11-13  & 20:51:35  & 6.7 & 5.45  & quiet & strike-slip  & 296  & weak response\\
\hline 
2015-11-24  & 22:45:40  & 7.6 & 18.01 & quiet & normal  & 165  & weak response\\
\hline 
2015-12-07  & 07:50:07  & 7.2 & 12.69 & AE disturb. & strike-slip  & 443  & weak response\\
\hline 
\end{tabular}
\caption{
Results of the GPS-TEC observations in the vicinity of earthquake epicenter as well as parameters of the earthquakes and space weather disturbance level in 2013-2015.
}
\label{tab:2} 

\end{table}

\newpage

\newpage
\begin{table}
\tiny
\begin{tabular}{|c|c|c|c|c|p{1cm}|c|c|}
\hline 
Date & $K_{W}$ & S.station & Period (sec) & LST & foF2 (MHz) & M & Effect\\
\hline 
\hline 
25/04/2015{*} & 8.8 & TATO & 13.7 & 14.6  & 14.5{*} & 7.8 & {*}m/h 250km, no effect\cite{Chum_2016}\\
\hline 
11/03/2011{*} & 8.0 & TATO & 22.7 & 14.0  & 11{*} & 9.0 & {*}Multicusp\cite{Liu_2011}\\
\hline 
11/03/2011	& 6.6	&TLY &	7.9	&	13.0& 9 & 9.0 & Multicusp \\
\hline
27/02/2010{*} & 5.8 & ARU & 11.3 & 11.2 & 7{*} & 8.8 & {*}Multicusp\cite{Maruyama_2016a}\\
\hline 
11/04/2012(a)	& 5.7	&TLY &	17.6	&	16.2& 8.5 & 8.4 & Multicusp\\
\hline 
24/09/2013	& 5.7	&TLY &	8.0	&	18.9 & 7 & 7.7 & no effect \\
\hline 
24/05/2013	& 5.6	&TLY &	25.2	&	13.0 & 7 & 8.3  & Multicusp \\
\hline 
25/04/2015	& 5.6	&TLY &	10.9	&	13.5&  9.5 & 7.8  & Multicusp\\
\hline 
17/09/2015	& 5.4	&TLY &	11.4	&	7.3& 4.2 & 7.0  & Multicusp \\
\hline 
16/04/2013	& 5.2	&TLY &	16.5	&	18.1 & 8.5 & 7.8  & Disturbed conditions \\
\hline 
11/04/2012(b)	& 5.1	&TLY &	11.2	&	18.3& 7.5 & 8.0 &  Multicusp\\
\hline 
25/10/2013	& 5.1	&TLY &	7.4	&	0.5 & 4.5 & 7.1  & Esp bifurcation\\
\hline 
12/05/2015	& 4.9	&TLY &	11.7	&	14.4 & 7 & 7.3  & Multicusp\\
\hline 
25/04/2015{*} & 4.8 & NKC & 17.2 & 07.8  & 7.5{*} & 7.8 & {*}Multicusp\cite{Chum_2016}\\
\hline 
12/04/2014	& 4.8	&TLY &	7.9	&	3.9& 4 & 7.6  & m/h 270km, no effect \\
\hline 
16/09/2015	& 4.8	&TLY &	8.5	&	6.3& 4 & 8.3  & Disturbed  conditions\\
\hline 
26/02/2012	& 4.7	&TLY &	7.3	&	13.3& 8.0 & 6.6 & Multicusp\\
\hline 
14/02/2013	& 4.7	&TLY &	41.9	&	20.5 & 4.2 & 6.6  & m/h 250km, no effect\\
\hline 
07/12/2015	& 4.6	&TLY &	3.2	&	15.1 & 7 & 7.2 & Esp bifurcation\\
\hline 
02/03/2016	& 4.5	&TLY &	9.9	&	20.4& 5.0 & 7.8 & no effect\\
\hline 
20/04/2013	& 4.4	&TLY &	7.1	&	7.3 & 7 & 6.6 & Disturbed conditions\\
\hline 
17/04/2016	& 4.4	&TLY &	10.7	&	8.3& 4 & 7.8 & Disturbed conditions, no effect \\
\hline 
07/12/2012	& 4.2	&TLY &	7.6	&	15.6& 7.0 & 7.3 & no effect\\
\hline 
28/09/2013	& 4.2	&TLY &	8.8	&	15.0 & 8 & 6.8  & no effect \\
\hline 
16/02/2015	& 4.2	&TLY &	6.0	&	6.4 & 3 & 6.8 & Esp bifurcation, m/h 250km\\
\hline 
26/10/2015	& 4.2	&TLY &	54.6	&	16.4 & 8 & 7.5  & no effect \\
\hline 
15/04/2016	& 4.2	&TLY &	53.1	&	23.7& 4 & 7.0 & Disturbed conditions\\
\hline 
29/07/2016	& 4.1	&TLY &	62.4	&	4.7& 3.5 & 7.7 & Disturbed conditions, no effect\\
\hline 
26/04/2015	& 3.6	&TLY &	5.0	&	14.4 & 9 & 6.7  & Multicusp\\
\hline 
13/11/2015	& 3.5	&TLY &	13.2	&	4.1 & 3.5 & 6.7  & no effect, m/h 250km\\
\hline 
20/04/2015	& 3.2	&TLY &	7.6	&	9.1 & 9.2 & 6.4  & no effect \\
\hline 
30/01/2016	& 3.1	&TLY &	273.4	&	10.7& 8.0 & 7.2 & Disturbed conditions\\
\hline 
\end{tabular}
\caption{
Earthquakes on Rayleigh efficiency classes $K_{W}$. Asterisks
denote effects according to other publications with the calculated
coefficients over the data from corresponding seismic stations in
the vicinity of observation point. 
m/h defines the minimal observed height of F-track at the ionogram.
(*) Asterisk marks the data, available from some other publications.
}
\label{tab:3} 

\end{table}


\end{document}